\documentclass[12pt,draftcls,journal,onecolumn]{IEEEtran}

\usepackage{graphicx, epsfig}

\begin{document}

\title{Minimum Mean-Square-Error Equalization using Priors for Two-Dimensional Intersymbol Interference}

\author{Naveen Singla and Joseph A. O'Sullivan\thanks{This work was supported in part by the Boeing Foundation and by the Office of Naval Research under Award N00014-03-1-0110.} 
\thanks{The authors are with the Electronic Systems and Signals Research Laboratory, Department of Electrical and Systems Engineering, Washington University in St. Louis, MO 63130, USA (e-mail: jao@ese.wustl.edu). This work was presented in part at the 2004 International Symposium on Information Theory~\cite{singla2}.}}

\maketitle

\begin{abstract}

Joint equalization and decoding schemes are described for two-dimensional intersymbol interference (ISI) channels. Equalization is performed using the minimum mean-square-error (MMSE) criterion. Low-density parity-check codes are used for error correction. The MMSE schemes are the extension of those proposed by T\"{u}chler \emph{et al.} (2002) for one-dimensional ISI channels. Extrinsic information transfer charts, density evolution, and bit-error rate versus signal-to-noise ratio curves are used to study the performance of the schemes.


\end{abstract}


\section{Introduction}

T\"{u}chler \emph{et al.}~\cite{tuchler1} considered the problem of coded data transmission over one-dimensional intersymbol interference (ISI) channels. Receivers based on the principle of turbo equalization have proven highly successful for these channels. Such a receiver consists of multiple equalizers/decoders that exchange extrinsic information and each component computes its output using the extrinsic information of the other components (as \emph{a priori} information) along with the channel output. T\"{u}chler \emph{et al.} proposed soft-in soft-out equalizers based on the minimum mean-square-error (MMSE) criterion that use the extrinsic information from the error-correction code decoder to compute their estimate. They showed that the performance of their iterative receiver, an MMSE equalizer with a convolutional code, is very close to that of the iterative receiver using the much more complex MAP (BCJR) equalizer. 

In this letter we consider coded data transmission over two-dimensional (2D) ISI channels. The absence of (computationally tractable) exact MAP or ML algorithms for two dimensions necessitates the search for low-complexity approximate schemes. Many such schemes have been proposed by Singla \emph{et al.}~\cite{singla1}, Marrow and Wolf~\cite{marrow}, and Shental \emph{et al.}~\cite{shental} among others. We describe iterative schemes that use two dimensional extensions of the aforementioned MMSE equalizers in conjunction with low-density parity-check codes for the 2D ISI. The equalizers are modified taking into account the 2D nature of the ISI.

MMSE equalization for 2D ISI channels was first proposed and studied by Chugg \emph{et al.}~\cite{chugg2}. However, the equalizer they proposed is not iterative and they did not employ any error-correction coding. Singla \emph{et al.}~\cite{singla1} also proposed an iterative receiver for 2D ISI channels using MMSE equalization followed by decoding using low-density parity-check (LDPC) codes. However, the MMSE equalizers proposed herein have a much lower complexity and better performance than those previously proposed by Singla \emph{et al.}~\cite{singla1}. The rest of the paper is organized as follows. In Section 2 describes the model we use for the systems with 2D ISI. In Section 3, we describe three MMSE equalization schemes using a linear equalizer, and iterative decoding algorithms using LDPC codes with the equalizers. Results are provided in Section 4 and conclusions in Section 5.

\section{System Model}


\noindent{The channel is represented as a discrete-time channel governed by the following equation}

\begin{equation}
r(i,j) = \sum_{l_1=0}^{L-1}\sum_{l_2=0}^{L-1} x(i-l_1,j-l_2)h(l_1,l_2) + w(i,j),
\label{eq:one}
\end{equation}


\noindent{where $r(i,j)$ and $x(i,j)$ are elements of the output and encoded data matrices, respectively; $w(i,j)$ are samples of the noise, assumed to be additive white Gaussian; ${\bf{h}}=\{h(l_1,l_2)\}_{l_1,l_2=0}^{L-1}$ is the 2D channel point spread function. For error correction we use LDPC coset codes~\cite{kavcic} whose code graph chosen uniformly at random from the ensemble of regular graphs. The ISI coefficients are assumed to be real. For the purpose of illustration of our concepts and simulations we use $3{\times}3$ point spread functions of the form}

\begin{equation}\label{eq:isi1}
{\bf{h}} = \left(\begin{array}{ccccc}
   c  &&  b  && c\\
   b  &&  a  && b\\
   c  &&  b  && c
\end{array} \right),
\end{equation}

\noindent{where the interference coefficients are specified in terms of an ``interference parameter'' $s$: $a=1/\sqrt{1+s^2+s^4}$, $b=a{\cdot}s/2$, and $c=a{\cdot}s^2/2$. The parameter $s$ quantifies the the amount of interference; the larger the $s$, the more severe the interference.





\section{MMSE Equalization and Decoding}




The equalizer computes the linear MMSE estimate of the data using the channel output and the extrinsic information from the LDPC decoder. The linear MMSE estimate of $x(i,j)$ is, 



\begin{equation}
\hat{x}(i,j) = E[x(i,j)]+\sum_{l_1=-N}^{N}\sum_{l_2=-N}^{N}\Big(r(i-l_1,j-l_2)-E[r(i-l_1,j-l_2)]\Big)c(i,j:l_1,l_2),
\label{eq:two}
\end{equation}

\noindent{where it is assumed that the estimate is computed using a $(2N+1){\times}(2N+1)$ support. The size of the support determines the complexity of the equalizer and there is usually a trade-off involved between performance and complexity of the equalizer. $\{c(i,j:l_1,l_2)\}_{l_1,l_2=-N}^{N}$ are the MMSE filter coefficients for bit $x(i,j)$ and $E[x(i,j)]$ is the mean of $x(i,j)$. The mean and variance of $x(i,j)$ can be calculated using the extrinsic information. The coefficients of the Wiener filter are obtained by solving the Wiener-Hopf equations}


\begin{equation}
{\bf{K_{rr}}}{\bf{c}}_{ij}={\bf{K}}_{{\bf{r}}x},
\label{eq:three}
\end{equation}

\noindent{where $\mathbf{K_{rr}}=E[(\mathbf{r}_{ij}-E[\mathbf{r}_{ij}])(\mathbf{r}_{ij}-E[\mathbf{r}_{ij}]^{T})]$; $\mathbf{K}_{\mathbf{r}{x}}=E[(\mathbf{r}_{ij}-E[\mathbf{r}_{ij}])(x(i,j)-E[x(i,j)])]$; $\mathbf{r}_{ij}$ is the $(2N+1)^2\times1$ vector obtained by rastering $\{r(i-l_1,j-l_2)\}_{l_1,l_2=-N}^{N}$; and $T$ denotes matrix transposition. Using~(\ref{eq:one}) and~(\ref{eq:three}) we can obtain the filter coefficients and then the MMSE estimate is obtained using~(\ref{eq:two}). During the computations for a particular bit its extrinsic information, obtained from the LDPC decoder, is set to zero which modifies the filter coefficients and the MMSE estimate. This is done to ensure that the estimate contains only extrinsic information.}


As in~\cite{tuchler1}, it is assumed that after MMSE equalization the probability density functions (pdf) $p(\hat{x}(i,j){\mid}x(i,j)=x), x{\in}\{{\pm1}\},$ are Gaussian with parameters $\mu_{ij}(x)=E[\hat{x}(i,j){\mid}x(i,j)=x]$ and $\sigma_{ij}^{2}(x)=Cov(\hat{x}(i,j),\hat{x}(i,j){\mid}x(i,j)=x)$. The statistics $\mu_{ij}(x)$ and $\sigma_{ij}^{2}(x)$ can be calculated using the filter coefficients. Under this assumption the extrinsic information (the log-likelihood ratio (LLR)) of the data, from the equalizer becomes

\begin{equation}
L_{E}(x(i,j))=\frac{2\hat{x}(i,j)\mu_{ij}(+1)}{\sigma_{ij}^{2}(+1)}.
\label{eq:thirteen}
\end{equation}

\noindent{The equalizer sends the extrinsic information to the LDPC decoder which uses it as \emph{a priori} information and performs a fixed number of sum-product message-passing iterations~\cite{rich} before passing its extrinsic information to the MMSE equalizer. This process is continued until the receiver converges or a maximum number of iterations is exhausted.}

For the scheme described above the computation of the filter coefficients involves inversion of a $(2N+1)^2{\times}(2N+1)^2$ matrix which causes a high computational load. One approximation which can be used to reduce this load is to have time-invariant coefficients. Following~\cite{tuchler1} we investigate the performance of the exact equalizer when the filter coefficients are calculated assuming no prior information and perfect prior information, referred to as the approximate I and approximate II schemes, respectively. In either case the filter coefficients are calculated only once and used for all the iterations. The extrinsic information is still calculated as in~(\ref{eq:thirteen}) with the MMSE estimate now being calculated using either of the approximate schemes. The approximate II equalizer turns out to be the matched filter implementation as noted by T\"{u}chler \emph{et al.}~\cite{tuchler1}.

\section{Results and Discussion}

\subsection{Extrinsic Information Transfer Charts}

We use extrinsic information transfer (EXIT) charts~\cite{tenbrink} to compare the performance of the equalizers described in the previous section. EXIT charts show how the ``quality'' of the output information varies with the quality of the input information for a particular receiver component. For the EXIT analysis, the equalizer is modeled as a device mapping the channel output $\mathbf{R}$ and the \emph{a priori} LLRs $L_i$ to a sequence of output LLRs $L_o$. It is assumed that the sequences $L_i$ and $L_o$ are independent and identically distributed Gaussian and that the magnitude of the mean of the Gaussians is equal to half the variance, thus the LLRs can be specified by a single parameter. Using ten Brink's approach, we plot $I_o$, the mutual information between $L_o$ and $X$, versus $I_i$, the mutual information between $L_i$ and $X$. Here $X$ is a binary valued random variable taking values +1 or -1 with equal probability. The pdf of $L_o$ is estimated by making a histogram of the LLR values at the equalizer output. 


Fig.~\ref{fig:exit} shows the exit charts for the equalization schemes for the point spread function corresponding to $s$=0.4. The SNR is calculated as

\begin{equation}
SNR=10{\cdot}{\log_{10}}\frac{\sum_{l_1,l_2=0}^{L-1}h^{2}(l_1,l_2)}{2\sigma_{w}^{2}}.
\label{eq:SNR}
\end{equation}

The filters each have a $5{\times}5$ support. For each EXIT chart, $10^6$ randomly chosen equiprobable symbols $x(i,j){\in}{\pm1}$ were generated and transmitted over the ISI channel. As expected the exact MMSE scheme has the best performance. The approximate I scheme has a good starting behavior, but poor behavior at high values of $I_i$, whilst the opposite is observed for the approximate II scheme, which is consistent with the observations for one-dimensional ISI channels. As observed in~\cite{tuchler1}, the EXIT charts for the two approximate schemes in Fig.~\ref{fig:exit} suggest using a scheme which switches between the two approximate equalizers based on which equalizer yields a larger value of $I_o$ for a given $I_i$. We investigate this scheme, termed the hybrid scheme, in the following subsections.

The figure also shows the EXIT charts for the exact MMSE equalizer at different SNRs. Also shown is the EXIT chart for a block length 10000, regular (3,6) LDPC code. As the SNR is reduced, the gap between the EXIT charts of the equalizer and decoder becomes narrower until the two touch at an SNR of about -1.05 dB. Reducing the SNR beyond this point leads to the decoding trajectory getting stuck at the point of intersection leading to poor performance. Thus this value of the SNR gives us an idea of how much noise the equalizer-decoder pair can tolerate so as to reliably recover the data. 






\subsection{Density Evolution Using Gaussian Approximation}{\label{gaussevol}}

EXIT charts have proven very useful in predicting the behavior of iterative decoders. However, the process of determining the noise threshold using EXIT charts is quite tedious. In this section we propose a density evolution algorithm to determine noise tolerance thresholds for the MMSE-LDPC schemes. The density evolution algorithm uses the Gaussian approximation that was used for the EXIT charts, \emph{i.e.}, the pdf of the input/output of the receiver components are Gaussian and can be characterized by a single parameter. This approximation was shown to be very accurate for the sum-product decoder for LDPC codes by Chung \emph{et al.}~\cite{chung}.

For the density evolution algorithm we evolve the mean of the densities through the iterations. The density evolution algorithm proceeds as follows: at every iteration the MMSE equalizer computes the mean of the output LLR and passes it to the LDPC decoder; using that mean the LDPC decoder computes the mean of its output LLR and passes it to the MMSE equalizer. If the mean tends to infinity as the iterations progress then the variance is increased and the same process is repeated. This continues until the critical value of variance when the mean does not tend to infinity. The density evolution for LDPC codes using a Gaussian approximation is described in~\cite{chung}. For the sake of brevity we omit the description.

Table~\ref{tbl:thresmmseldpc3} lists the thresholds for the MMSE-LDPC schemes for the $3{\times}3$ point spread functions defined by~(\ref{eq:isi1}). The SNR is calculated as in~(\ref{eq:SNR}) except that the rate of the LDPC code is also taken into account. The LDPC code is the regular (3,6) code. The table shows that as the interference becomes more severe the thresholds become worse, which is what we expect. The table shows that the noise thresholds of the hybrid scheme are very close to those of the exact scheme. The thresholds for the approximate II scheme are not shown since they are very high.




\subsection{Bit-Error Rate versus SNR Curves}

Fig.~\ref{fig:snrber1} shows the bit-error rate (BER) versus SNR curves for the MMSE-LDPC schemes, with SNR defined in~(\ref{eq:SNR}). The performance of the MMSE-LDPC schemes is plotted for the different equalizers for the ISI in~(\ref{eq:isi1}) corresponding to $s$=0.4. A block length 10000, regular (3,6) LDPC code is used and the leftmost curve shows the performance of this code on an AWGN channel. The figure also shows the performance of the full graph algorithm, a sum-product message-passing based receiver for the 2D ISI channel~\cite{singla1}. For the MMSE-LDPC schemes the equalizer performs a maximum of ten iterations; 20 iterations of LDPC decoding are performed for each equalization. The BER curves confirm what the noise thresholds suggested, namely that the hybrid scheme's performance is very close to that of the exact scheme. The approximate I scheme suffers a loss of about 1.5 dB compared to the exact scheme. Again the performance of the approximate II scheme is omitted since it is very bad.







Fig.~\ref{fig:snrber1} also shows the performance for the exact MMSE-LDPC scheme for increasing block lengths. The dashed vertical line depicts the threshold for the exact MMSE-LDPC scheme using a regular (3,6) code as calculated in the previous subsection. It can be seen from the figure that, even as the block length increases, very low BERs are achieved only when the SNR is above the threshold SNR. Fig.~\ref{fig:snrber2} shows the performance curves for the exact MMSE-LDPC scheme for ISI corresponding to different values of $s$ in~(\ref{eq:isi1}). As expected, when the ISI becomes severe the performance of the receiver degrades. However, even for $s=0.6$, when the interference energy is nearly 33\%, the loss in SNR over the LDPC code is only a little over 3 dB.  
\section{Conclusions}

In this letter we have presented results for the application of the MMSE schemes proposed by T\"{u}chler \emph{et al.}~\cite{tuchler1} to 2D ISI channels. The MMSE equalizers are used in conjunction with LDPC decoding to further improve the performance. Simulations results based on EXIT charts, density evolution, and BER versus SNR curves, show similar performance trends as for one-dimensional ISI channels. The performance of the MMSE-LDPC receiver using the exact equalizer is very close to that of the message-passing based algorithm and at a much lower computational cost. The hybrid scheme performs almost as-well-as the exact scheme and has even lower computational complexity.


\section{Acknowledgments}

The authors would like to thank Andrew Singer and Robert Drost of the University of Illinois for helpful discussions about the equalization schemes and the interpretation and verification of the results.

\newpage

\begin{figure}[ht]
 \centering
  \scalebox{1.0}{\includegraphics{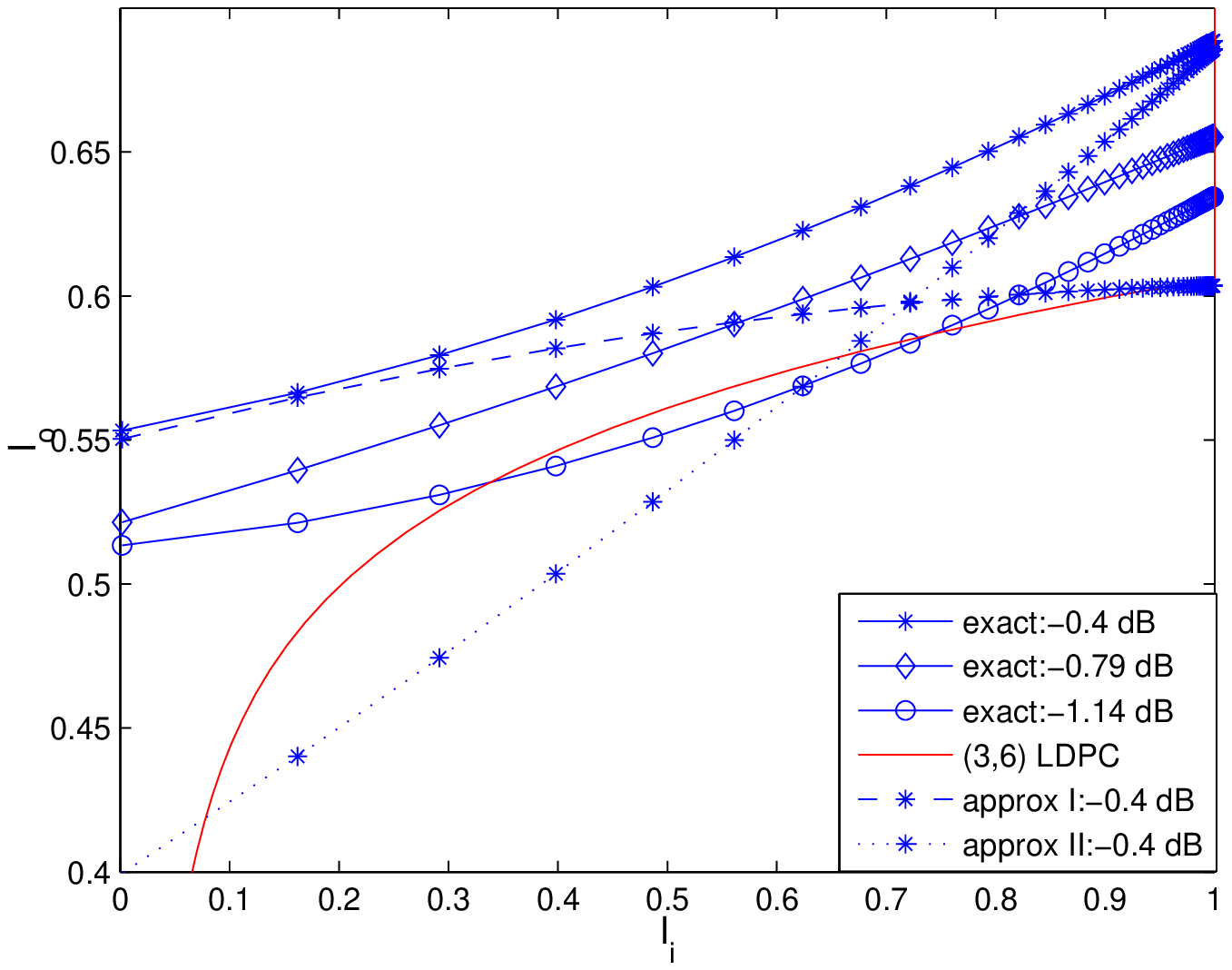}}
  \caption{ \label{fig:exit} EXIT charts for the MMSE equalization schemes at 1.15 dB SNR. EXIT charts for the exact MMSE scheme for different SNRs. Also shown is the EXIT chart for a block length 10000, regular (3,6) LDPC code.}
\end{figure}

\newpage

\begin{table}[ht]
\begin{center}
\caption{Thresholds for exact MMSE-LDPC decoding scheme using regular (3,6) LDPC code for different $3{\times}3$ point spread functions.}
\label{tbl:thresmmseldpc3}
\begin{tabular}{|c|c|c|c|c|}
\hline
s & Interference & \multicolumn{3}{|c|}{Threshold  SNR [dB]} \\ \cline{3-5}
  & Energy (\%)  & Exact  & Hybrid & Approx I \\
\hline
0.1 & 1.00  & 1.209 & 1.209 & 1.220 \\
\hline
0.2 & 3.99  & 1.349 & 1.385 & 1.445 \\
\hline
0.3 & 8.93  & 1.707 & 1.772 & 1.891 \\
\hline
0.4 & 15.65 & 2.197 & 2.247 & 2.472 \\
\hline
0.5 & 23.81 & 2.993 & 3.125 & 3.420 \\
\hline
0.6 & 32.87 & 4.168 & 4.306 & 4.622 \\
\hline
0.7 & 42.20 & 5.638 & 5.850 & 6.144 \\
\hline
0.8 & 51.21 & 7.594 & 7.932 & 8.125 \\
\hline
\end{tabular}
\end{center}
\end{table}

\newpage

\begin{figure}[ht]
 \centering
  \scalebox{1.0}{\includegraphics{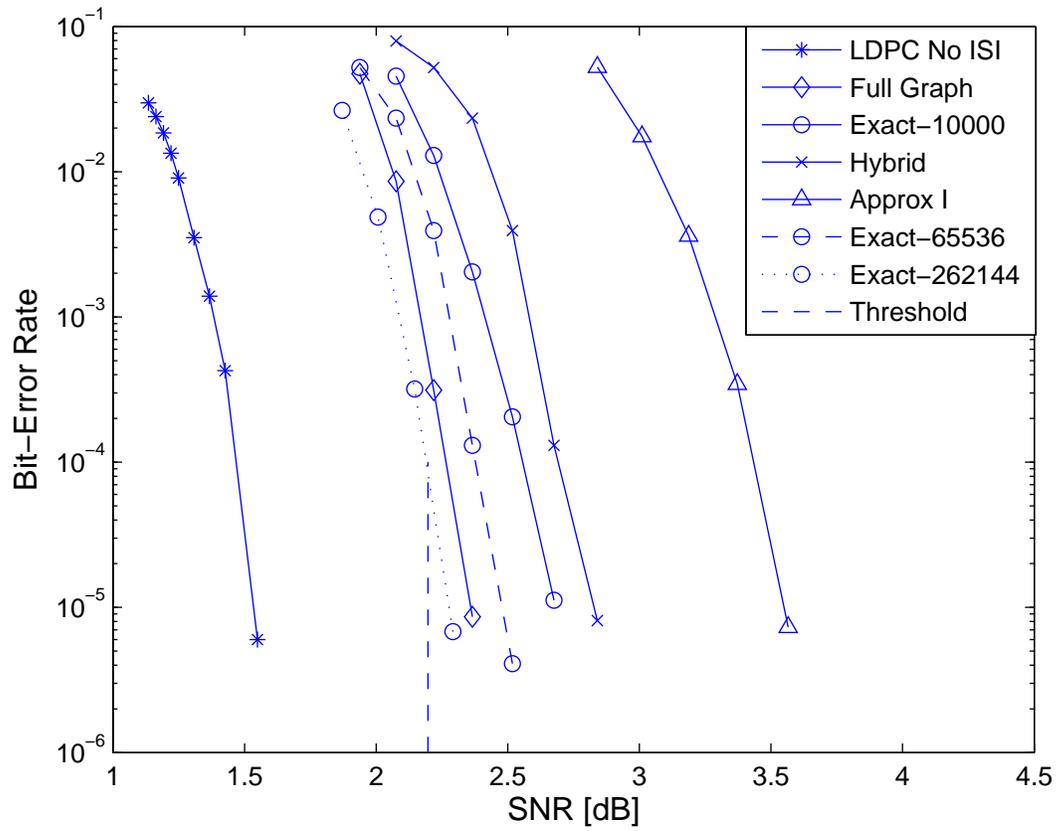}}
  \caption{ \label{fig:snrber1} BER versus SNR curves for different MMSE-LDPC schemes using a block length 10000, regular (3,6) LDPC code. BER versus SNR curves for the exact MMSE-LDPC scheme using regular (3,6) LDPC codes of increasing block lengths.}
\end{figure}

\newpage

\begin{figure}[ht]
 \centering
  \scalebox{1.0}{\includegraphics{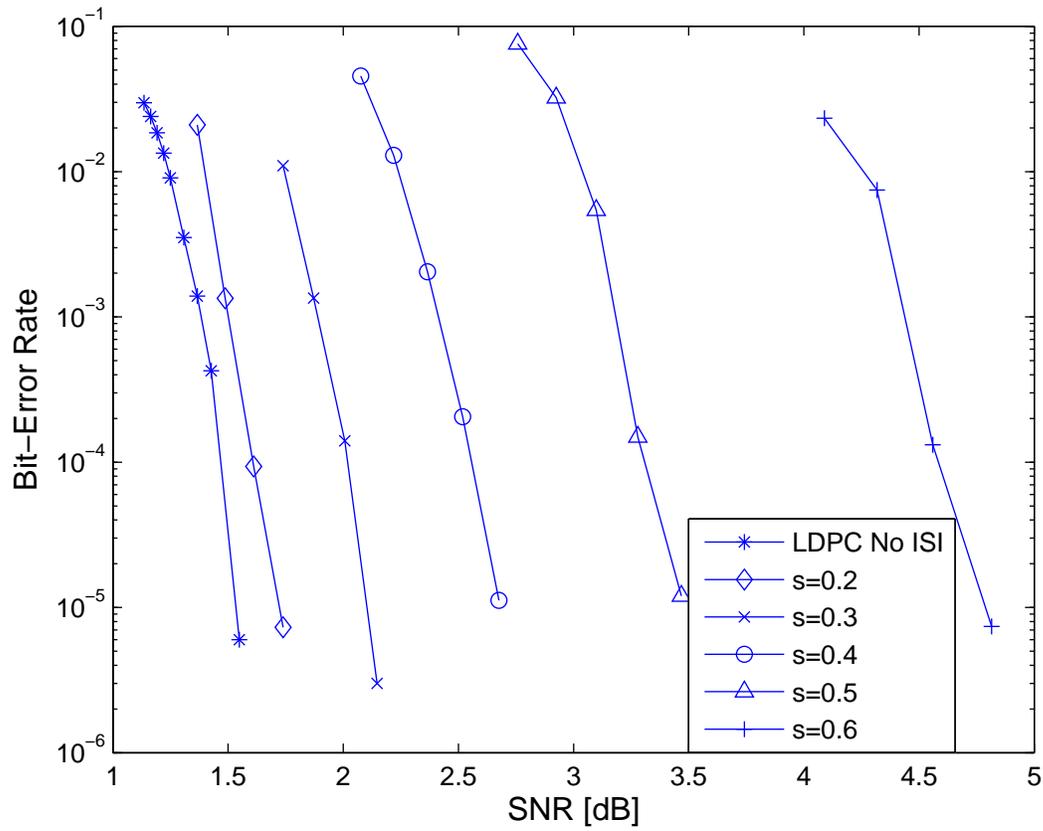}}
  \caption{ \label{fig:snrber2} BER versus SNR curves for the exact MMSE-LDPC schemes for different $3{\times}3$ point spread functions using a block length 10000, regular (3,6) LDPC code.}
\end{figure}

\end{document}